\documentclass[epj,twocolumn]{webofc}
\usepackage[varg]{txfonts}   
\usepackage{multirow}
\woctitle{Flavour changing and conserving processes}
\begin{document}
\title{Review of Recent Calculations of the Hadronic Vacuum Polarisation Contribution}
%
%

\author{Zhiqing Zhang\inst{1}\fnsep\thanks{\email{zhang@lal.in2p3.fr}} }

\institute{Laboratoire de l'Acc\'el\'erateur Lin\'eaire, Univ.\ Paris-Sud, CNRS/IN2P3, Universit\'e Paris-Saclay, Orsay, France}

\abstract{%
Recent calculations of the hadronic vacuum polarisation contribution are reviewed. The focus is put on the leading-order contribution to the muon magnetic anomaly involving $e^+e^-$ annihilation cross section data as input to a dispersion relation approach. Alternative calculation including tau data is also discussed. The $\tau$ data are corrected for various isospin-breaking sources which are explicitly shown source by source.
}
\maketitle
\section{Introduction}
\label{intro}
\begin{figure}[htb]
\centering
\includegraphics[width=\columnwidth,clip]{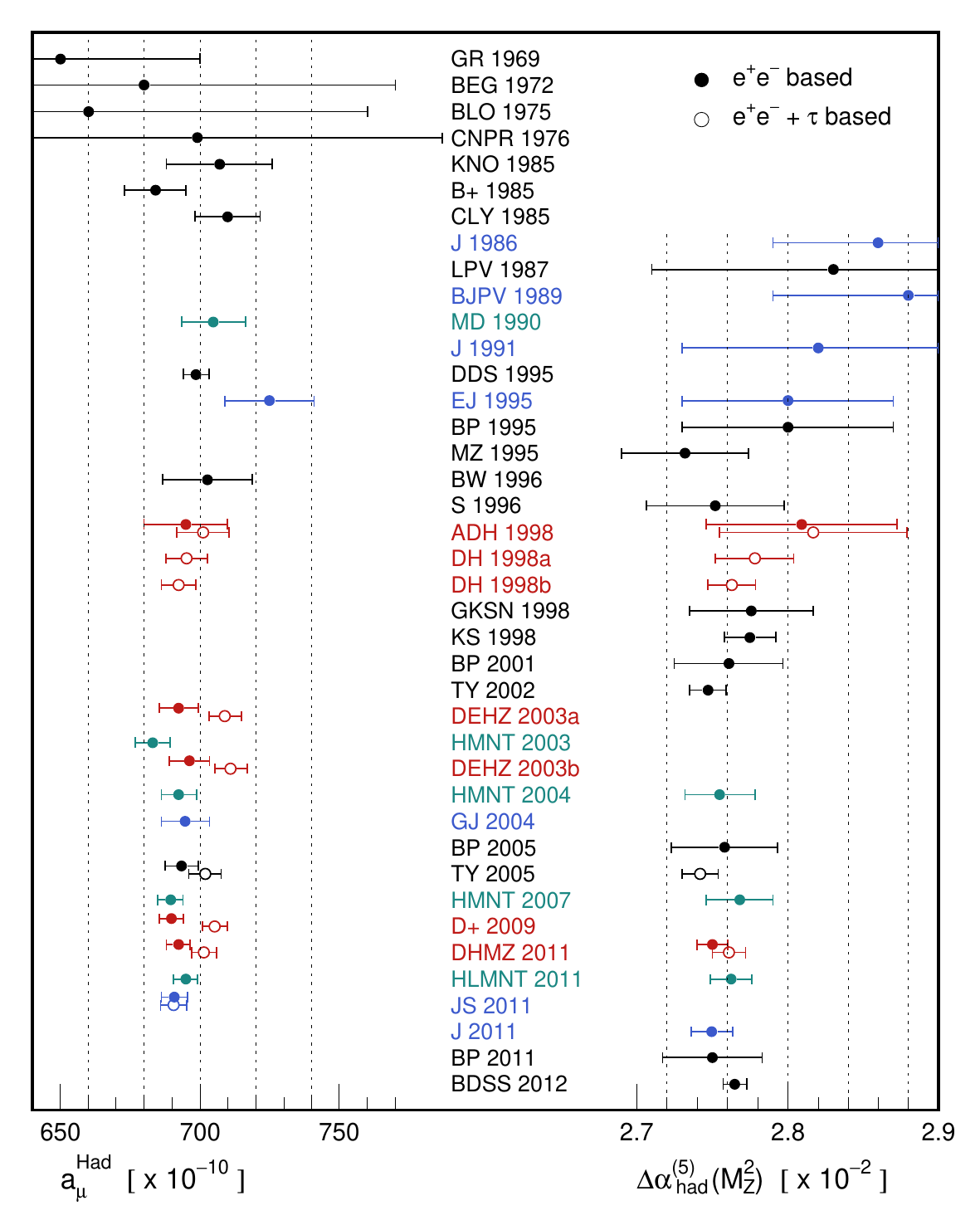}
\caption{Summary of the evolution as function of time for $a_\mu^{\rm Had}$ and $\Delta\alpha_{\rm had}^{(5)}(M^2_Z)$ (figure taken from \cite{cern60})}
\label{fig:evolution}
\end{figure}

The hadronic vacuum polarisation (HVP) contribution to the muon magnetic anomaly $a_\mu\equiv (g-2)/2$ may be decomposed into three parts as $a_\mu^{\rm Had}=a_\mu^{\rm Had, LO}+a_\mu^{\rm Had, HO}+a_\mu^{\rm had, LBL}$ corresponding to the leading-order (LO), higher-order (HO) and light-by-light (LBL) scattering contribution, respectively. The corresponding representative numerical values are $692.3\pm 4.2$~\cite{1010.4180}, $-9.79\pm 0.09$~\cite{1105.3149} and $10.5\pm 2.6$~\cite{0901.0306}, in units of $10^{-10}$ (the same units will implicitly be used for all following quoted $a_\mu$ numbers). Therefore the LO term is the dominant hadronic contribution and has the largest uncertainty not only among the three hadronic terms but also among all the electromagnetic, weak and hadronic sectors. This is why the improvement of the uncertainty of the LO HVP has been one of the main research activities by a number of groups and individuals since several decades (Fig.~\ref{fig:evolution}).

The HVP involving strongly interacting particles can be computed at large energy scales but not at low scales due to the non-perturbative nature of QCD at large distance. It is possible to overcome this problem by means of a dispersion relation technique involving experimental data on the cross section for $e^+e^-$ annihilation into hadrons. This will be the focus of this writeup. For alternative evaluations such as model dependent or Lattice QCD based calculations, we refer to \cite{maurice} and  \cite{lattice}. However the precision of the current Lattice calculation is still far from competitive with that of the dispersion relation approach.

This writeup is organised as follows. In Sec.~\ref{sec:amuee}, the state-of-the-art techniques used for the $e^+e^-$ based $a_\mu^{\rm Had,LO}$ calculation are reviewed. In Sec.~\ref{sec:amutau}, we discuss an alternative evaluation by including $\tau$ data and taking into account the known isospin-breaking corrections, followed by a summary in Sec.~\ref{sec:summary}.

\section{Review of the $\pmb{e^+e^-}$ based $\pmb{a_\mu^{\rm Had}}$ calculations}\label{sec:amuee}
The LO HVP contribution to $a_\mu^{\rm had}$ is calculated using the dispersion relation~\cite{dispersion} as
\begin{equation}
a_\mu^{\rm Had,LO}=\frac{1}{3}\left(\frac{\alpha}{\pi}\right)^2\int^\infty_{m^2_\pi} ds\frac{K(s)}{s}R^{(0)}(s)\,,
\end{equation}
where $K(s)$ is  a QED kernel function~\cite{kernel} and $R^{(0)}(s)$ represents the ratio of the bare cross section for $e^+e^-$ annihilation into hadrons to the point-like muon-pair cross section for a given centre-of-mass energy $\sqrt{s}$. The function $K(s)\sim 1/s$ gives a strong weight to the low energy part of the integral. Therefore, $a_\mu^{\rm Had, LO}$ is dominated by the $\rho(770)\to 2\pi$ resonance.
 
\begin{figure}
\centering
\includegraphics[width=\columnwidth,clip]{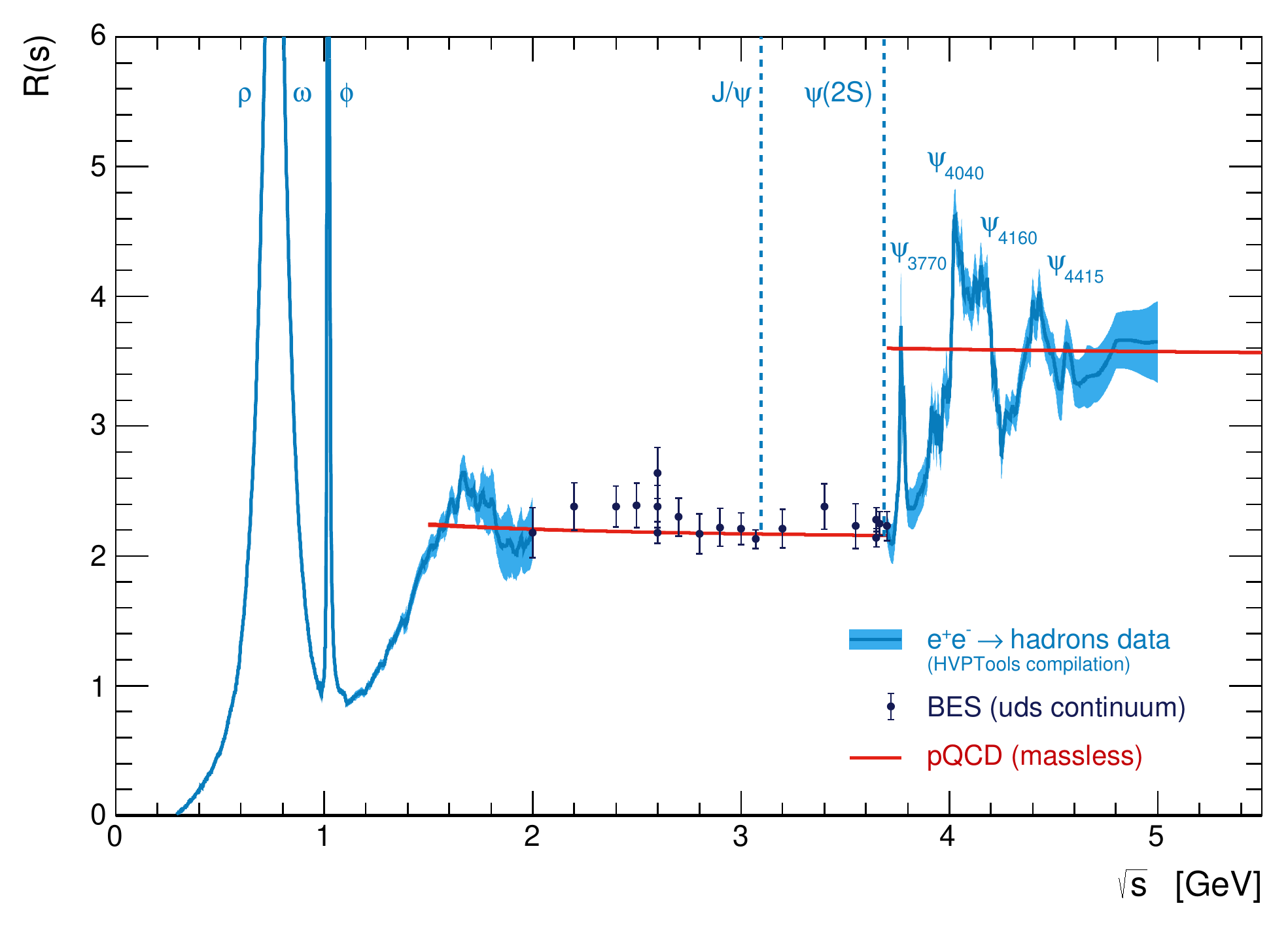}
\caption{Cross section for the process $e^+e^-\to$ hadrons versus centre-of-mass energy $\sqrt{s}$. The blue band represents the combined experimental measurements with their uncertainty. The red line shows the perturbative QCD prediction, the data points show the inclusive measurements from the BES experiment~\cite{bes} (figure taken from \cite{cern60})}
\label{fig:eeinput}
\end{figure}

The computation of the dispersion integral requires the knowledge of $R(s)$ at any scale $s$. Figure~\ref{fig:eeinput} shows an overview of the input data used in the evaluation of \cite{1010.4180}. 
In the low energy range between the $\pi^0\gamma$ threshold and 1.8\,GeV, the evaluation uses about 22 exclusive channels that are measured, and a few other channels that are (partially) estimated using isospin relations. In the continuum region between 1.8\,GeV and 3.7\,GeV, the prediction based on four-loop perturbative QCD calculation is used, which is in good agreement with the results from direct inclusive measurements by BES~\cite{bes}. The same QCD prediction is also used at higher energy scale above 5\,GeV. In the region just above the heavy quark thresholds and below 5\,GeV, the evaluation uses also $e^+e^-$ annihilation data to hadrons.

The evaluation of \cite{hlmnt11} uses a similar strategy. But there are a number of differences. For instance, the perturbative QCD predictions are only used between 2.6 and 3.73\,GeV and above 11.09\,GeV. The data treatment is also different. The evaluation of \cite{1010.4180} is based on HVPTools~\cite{dhmyz09} whereas that of~\cite{hlmnt11} using a clustering technique. A simplified comparison is shown in Table~\ref{tab:comp}.
\begin{table*}[htbp]
\centering
\caption{Comparison of data interpolation and combination between \cite{dhmyz09} and \cite{hlmnt11} where ISR, VP, FSR and $n_{\rm dof}$ stand for initial state radiation, vacuum polarisation, final state radiation correction and the number of degrees of freedom, respectively}
\label{tab:comp}
\begin{tabular}{lll}
\hline
 & HVPTools~\cite{dhmyz09} & Clustering method~\cite{hlmnt11}  \\\hline\noalign{\smallskip}
\multirow{2}{*}{Common points} & \multicolumn{2}{l}{Use bare cross sections (remove ISR, VP and add FSR to some early data)}\\
  & \multicolumn{2}{l}{Combine different experiments in a same channel before integration} \\\noalign{\smallskip}\hline\noalign{\smallskip}
Data interpolation & Using 2nd order polynomials & Trapezoidal rule (linear interpolation) \\ \noalign{\smallskip} \hline\noalign{\smallskip}
Normalisation & \begin{minipage}{6cm} Pseudo-MC generation fluctuates a data point along the original measurement taking into account correlation \end{minipage} & \begin{minipage}{6cm} Use directly measured data allowing the normalisation to float within quoted uncertainty \end{minipage}\\ \noalign{\smallskip} \hline \noalign{\smallskip}
Bin size & \begin{minipage}{6cm} Using small bins (1\,MeV) covered by the data point \end{minipage} & \begin{minipage}{6cm} Using varying bin/cluster size depending on data density \end{minipage} \\ \noalign{\smallskip} \hline \noalign{\smallskip}
Average &  \begin{minipage}{6cm}  Weighted average and covariance matrix calculated in the small bins \end{minipage} & \begin{minipage}{6cm} Average obtained with non-linear $\chi^2$ minimisation \end{minipage} \\ \noalign{\smallskip} \hline \noalign{\smallskip}
Error inflation & \multicolumn{2}{l}{Error scaled locally in a bin with $\sqrt{\chi^2/{n_{\rm dof}}}$ if it is greater than 1}\\ \noalign{\smallskip} \hline \noalign{\smallskip}
\end{tabular}
\end{table*}

It is instructive to look at Table~\ref{tab:comp1} (Table 4 from~\cite{hlmnt11}). The difference between the two evaluations are often larger than or comparable with the quoted uncertainties. There are also important difference on the quoted uncertainties. These differences could be a reflection of the different data and uncertainty treatment mentioned earlier. The use of different inputs in particular the old data sets could be another source though these old data sets are often imprecise and therefore their relative weight in the combination should be in general small. It is important that these differences could be better understood and reduced.
\begin{table*}[htbp]
\centering
\caption{Contributions to $a_\mu$ in the
  energy region from $0.305$ to $1.8$\,GeV from exclusive channels (Table 4 from~\cite{hlmnt11})}
\label{tab:comp1}
\begin{tabular}{c|c|c|c}
\hline
Channel & HLMNT (11)~\cite{hlmnt11} & DHMZ (10)\ \cite{1010.4180} & Difference \\
\hline
$\eta\pi^+\pi^-$   & $  0.88\pm0.10$ & $  1.15\pm0.19$ & $-0.27$ \\ 
$K^+K^-$           & $ 22.09\pm0.46$ & $ 21.63\pm0.73$ & $0.46$ \\
$K^0_SK^0_L$       & $ 13.32\pm0.16$ & $ 12.96\pm0.39$ & $ 0.36$ \\ 
$\omega\pi^0$      & $  0.76\pm0.03$ & $  0.89\pm0.07$ & $-0.13$ \\ 
$\pi^+\pi^-$       & $505.65\pm3.09$ & $507.80\pm2.84$ & $-2.15$ \\ 
$2\pi^+2\pi^-$     & $ 13.50\pm0.44$ & $ 13.35\pm0.53$ & $ 0.15$ \\ 
$3\pi^+3\pi^-$     & $  0.11\pm0.01$ & $  0.12\pm0.01$ & $-0.01$ \\ 
$\pi^+\pi^-\pi^0$  & $ 47.38\pm0.99$ & $ 46.00\pm1.48$ & $ 1.38$ \\ 
$\pi^+\pi^-2\pi^0$ & $ 18.62\pm1.15$ & $ 18.01\pm1.24$ & $ 0.61$ \\ 
$\pi^0\gamma$      & $  4.54\pm0.14$ & $  4.42\pm0.19$ & $ 0.12$ \\ 
$\eta\gamma$       & $  0.69\pm0.02$ & $  0.64\pm0.02$ & $ 0.05$ \\ 
$\eta2\pi^+2\pi^-$ & $  0.02\pm0.00$ & $  0.02\pm0.01$ & $ 0.00$ \\ 
$\eta\omega$       & $  0.38\pm0.06$ & $  0.47\pm0.06$ & $-0.09$ \\ 
$\eta\phi$         & $  0.33\pm0.03$ & $  0.36\pm0.03$ & $-0.03$ \\ 
$\phi(\rightarrow{\rm unaccounted})$
                   & $  0.04\pm0.04$ & $  0.05\pm0.00$ & $-0.01$ \\ 
\hline
Sum of isospin channels 
                   & $  5.98\pm0.42$ & $  6.06\pm0.46$ & $-0.08$ \\ 
\hline
Total              & $634.28\pm3.53$ & $633.93\pm3.61$ & $ 0.35$ \\ 
\hline
\end{tabular}
\end{table*}

\begin{figure}[htb]
\centering
\includegraphics[width=\columnwidth,clip]{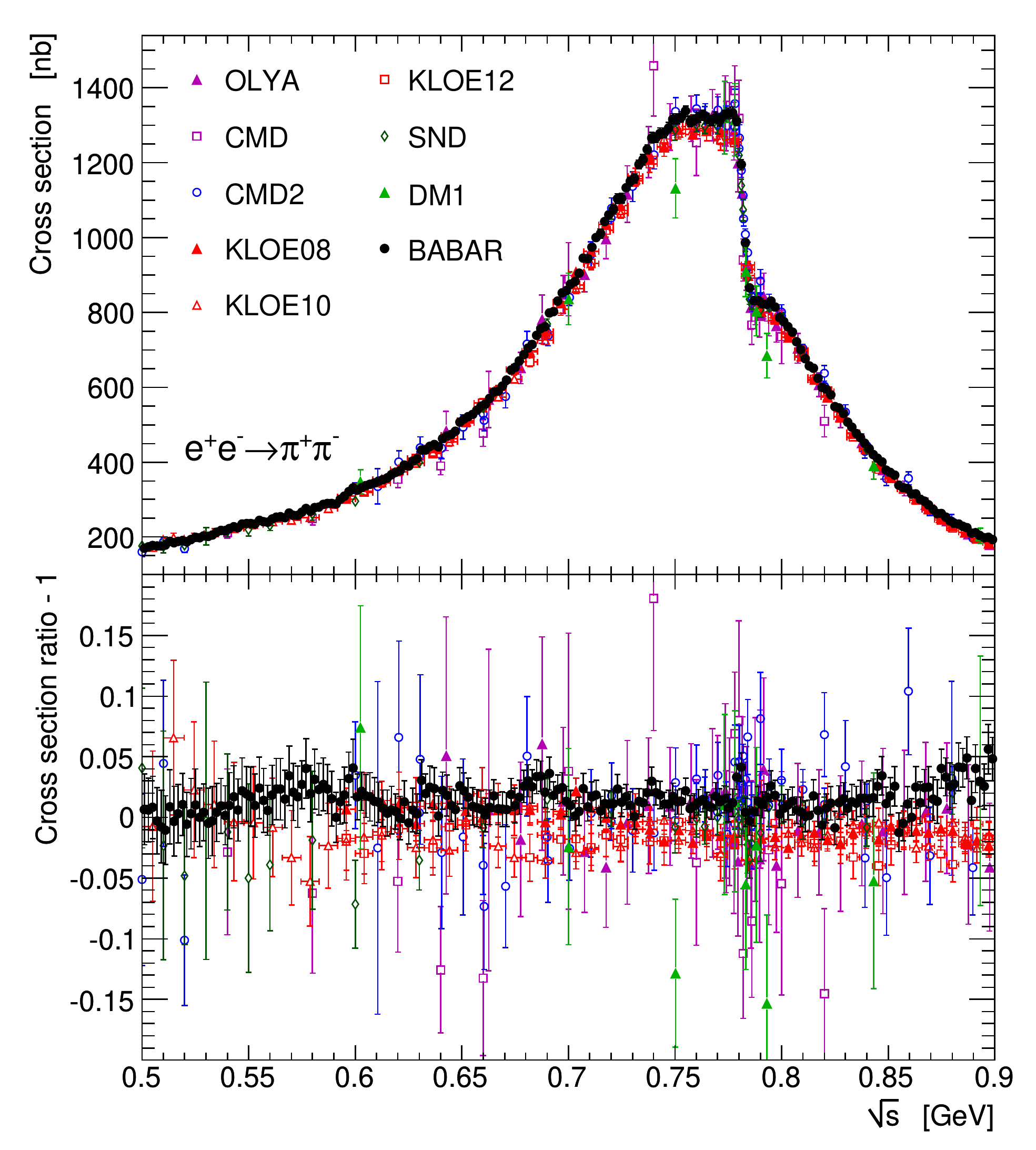}
\caption{Cross section for the process $e^+e^-\to \pi^+\pi^-$ measured by different experiments (top) and their relative difference with respect to the combined cross section (bottom). The uncertainty bars contain both the statistical and systematic components added in quadrature (figure taken from~\cite{cern60})}
\label{fig:2pi}
\end{figure}

\begin{figure*}[htbp]
\centering
\includegraphics[width=\columnwidth]{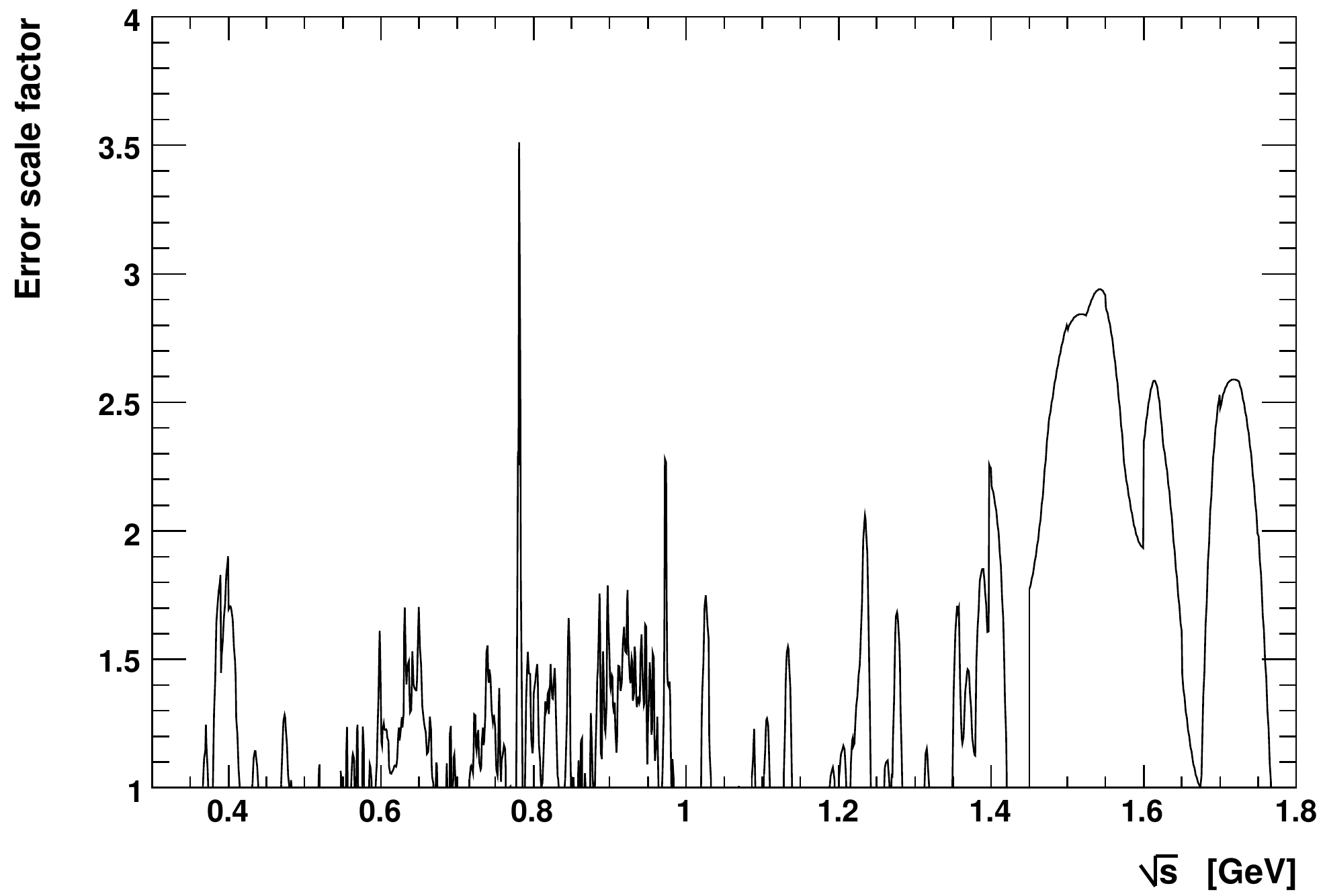}
\includegraphics[width=\columnwidth]{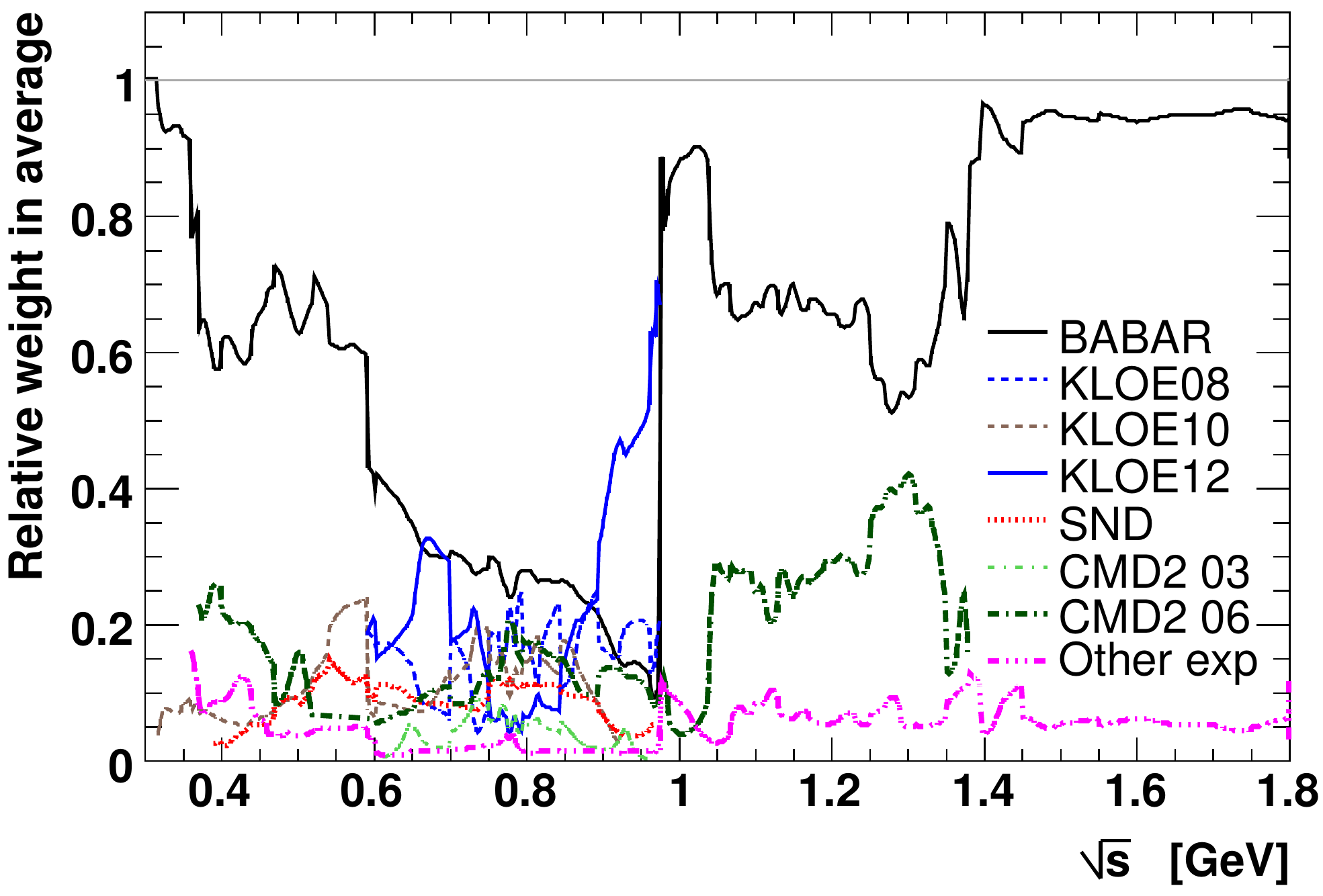}
\caption{Rescaling factor accounting for inconsistencies among experiments (left, Fig.~1 from \cite{dhmyz09}) and relative local averaging weight per experiment (right, Fig.~3 from \cite{1010.4180}) versus $\sqrt{s}$ in $e^+e^-\to \pi^+\pi^-$}
\label{fig:2pimore}
\end{figure*}

Table~\ref{tab:comp1} confirms that the $\pi^+\pi^-$ channel is by far the dominant one both in the central value and in the uncertainty. The corresponding inputs measured by different experiments are shown in Fig.~\ref{fig:2pi}. The inconsistency between Babar and KLOE in particular is clearly visible. Such inconsistency results in sizeable $\chi^2/{n_{\rm dof}}$ values (Fig.~\ref{fig:2pimore} (left)) preventing currently the expected precision improvement in the combination~\cite{dhmyz09}. The relative weight of each experiment is shown in Fig.~\ref{fig:2pimore} (right).

\begin{figure*}[htbp]
\centering
\includegraphics[width=\columnwidth]{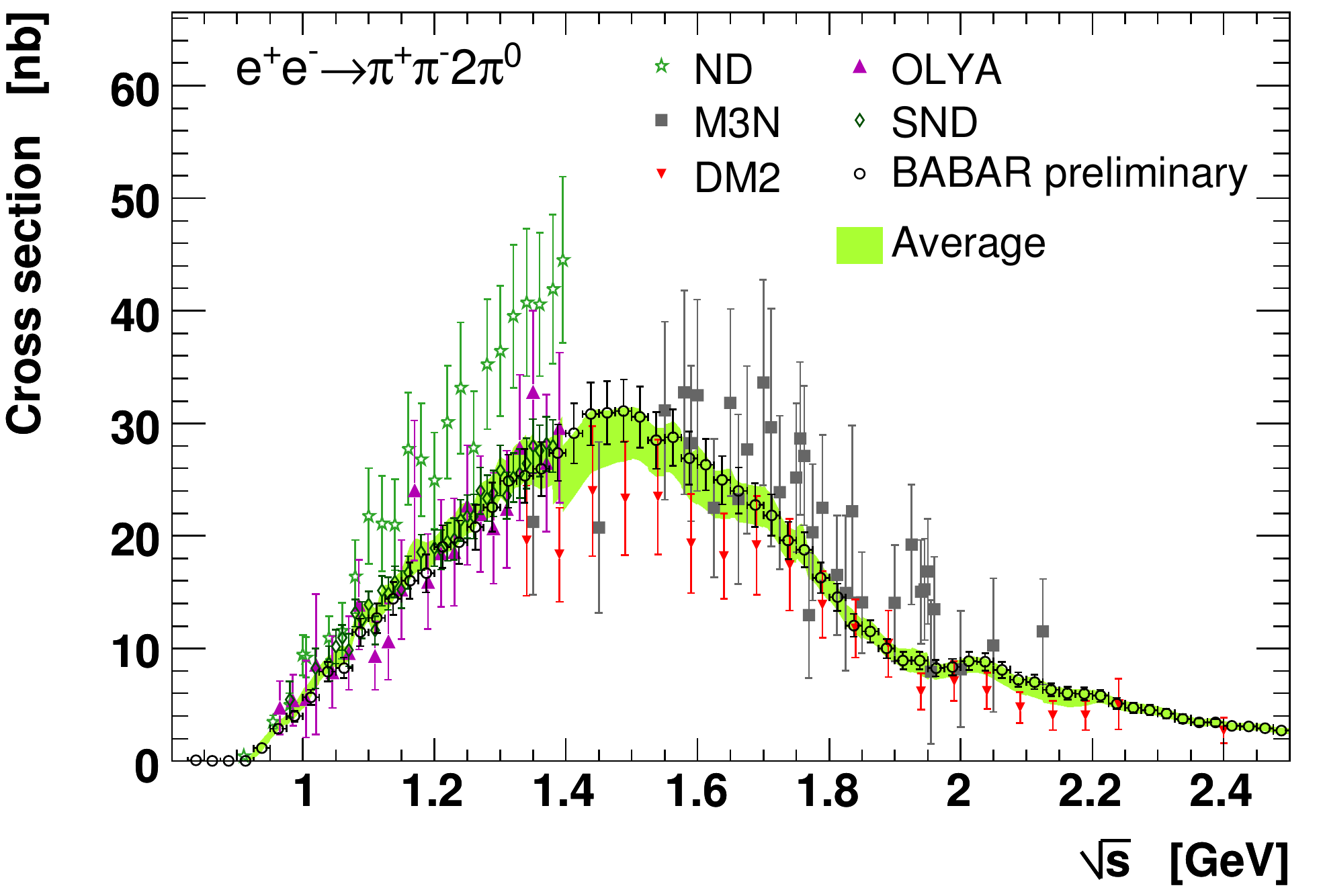}
\includegraphics[width=\columnwidth]{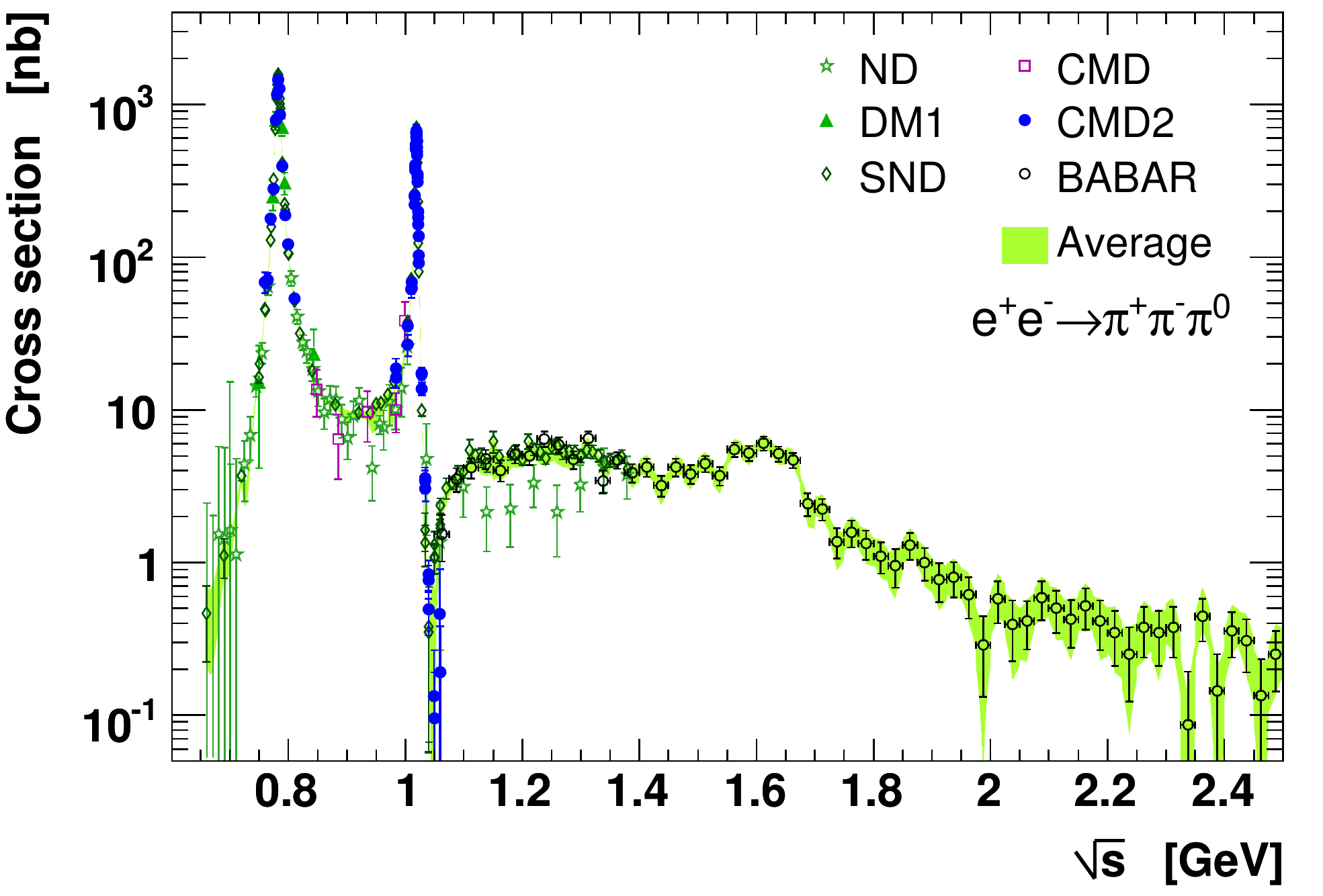}
\includegraphics[width=\columnwidth]{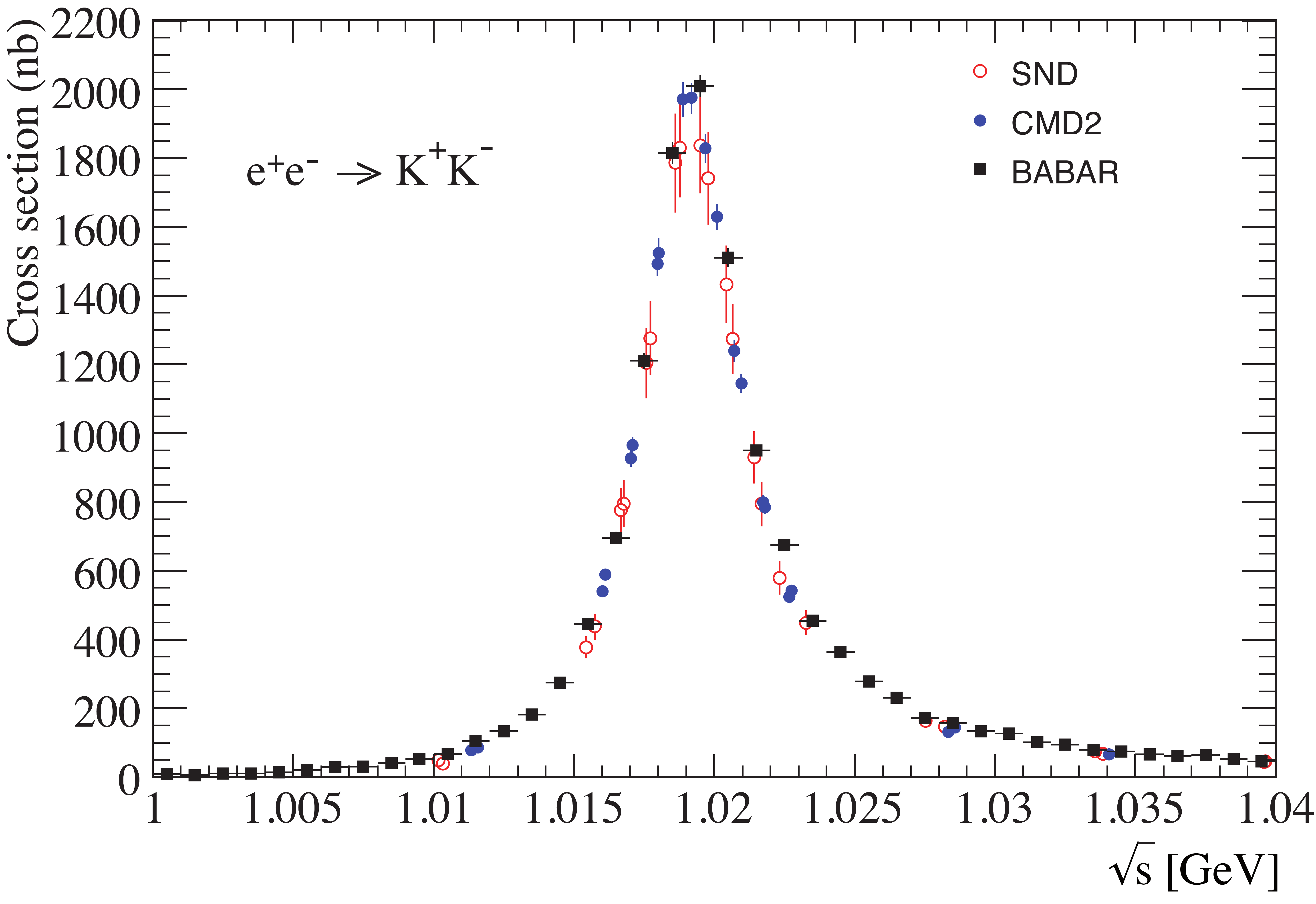}
\includegraphics[width=\columnwidth]{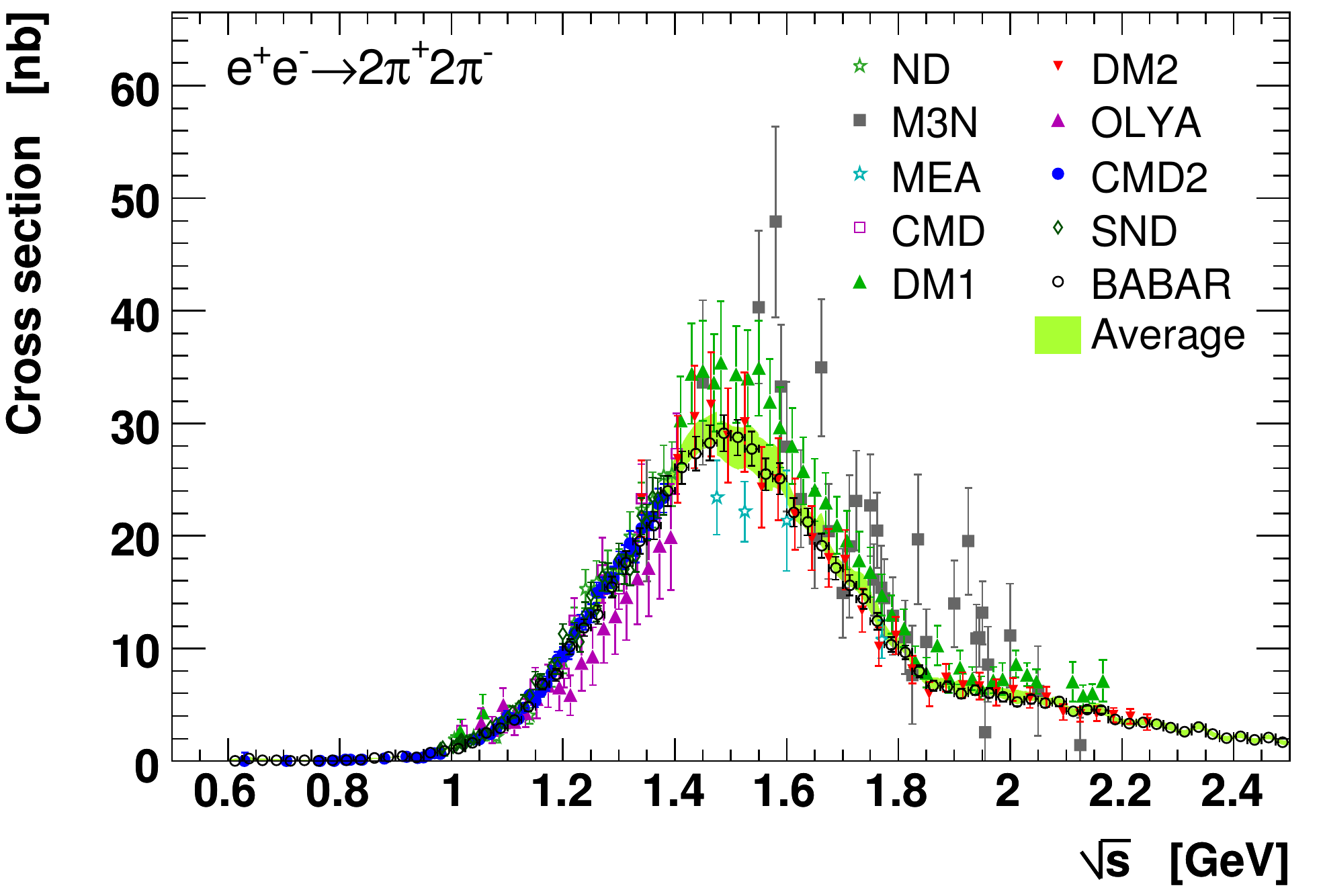}
\caption{Cross section inputs for the subleading channels $\pi^+\pi^-2\pi^0$ (top left), $\pi^+\pi^-\pi^0$ (top right), $K^+K^-$ (bottom left) and $2\pi^+2\pi^-$ (bottom right). The figures with the green error bands are taken from \cite{1010.4180} and the figure for the $K^+K^-(\gamma)$ channel from \cite{babarkk}. The error bands correspond to the $1\sigma$ combination uncertainty obtained with the HVPTools}
\label{fig:sub}
\end{figure*}

Table~\ref{tab:comp1} also shows that $\pi^+\pi^-2\pi^0$, $\pi^+\pi^-\pi^0$, $K^+K^-$ and $2\pi^+2\pi^-$ are important subleading channels contributing to the uncertainty budget. The corresponding measurements are shown in Fig.~\ref{fig:sub}. For the first channel, the Babar measurements dominate,  which are, however, still preliminary and the final results are expected soon and hopefully will improve the precision of the channel. For the $\pi^+\pi^-\pi^0$, the measurements in the $\omega$ and $\phi$ resonances are from the energy-scan experiments whereas the measurements above the resonances come almost exclusively from Babar. There is clearly room for improvement for this channel. The most precise measurements for the $\phi$ resonance in the $K^+K^-$ channel used to be from CMD-2 experiment. A new measurement from Babar has recently been published~\cite{babarkk}, which contributes to $a_\mu$ with $22.93\pm 0.28$ in the energy range from threshold to 1.8\,GeV, to be compared with the values in Table~\ref{tab:comp1}. There are many measurements for the $2\pi^+2\pi^-$ channel with the new Babar measurements~\cite{babar4pi} dominating in the precision of the combination. The updated result is $13.64\pm 0.36$, which improves the numbers quoted in Table~\ref{tab:comp1}.

\section{Alternative calculation of $\pmb{a_\mu^{\rm Had}}$}\label{sec:amutau}
In Fig.~\ref{fig:evolution}, some of the predictions are $e^+e^- + \tau$ based. The use of tau data of semi-leptonic $\tau$ decays in the evaluation of $a_\mu^{\rm Had}$ and $\Delta\alpha_{\rm had}^{(5)}$ was originally proposed in Ref.~\cite{adh98}. It is based on the fact that in the limits of isospin invariance, the spectral function of the vector current decay $\tau^-\to X^-\nu_\tau$ is related to the $e^+e^-\to X^0$ cross section of the corresponding isovector final state $X^0$ (so-called the conserved vector current (CVC) relation),
\begin{equation}
\sigma^{l=1}_{X^0}(s)=\frac{4\pi\alpha^2}{s}v_{1, X^-}(s)\,,
\end{equation}
where $s$ is the centre-of-mass energy-squared or equivalently the invariant mass-squared of the $\tau$ final state $X$, $\alpha$ is the electromagnetic fine structure constant, and $v_{1,X^-}$ is the non-strange, isospin-one vector spectral function given by
\begin{eqnarray}
v_{1,X^-}(s)&\!\!\!=\!\!\!& \frac{m^2_\tau}{6|V_{ud}|^2}\frac{{\cal B}_{X^-}}{{\cal B}_e}\frac{1}{N_X}\frac{dN_X}{ds}\nonumber \\
&\!\!\!\times\!\!\!& \left(1-\frac{s}{m^2_\tau}\right)^2\left(1+\frac{2s}{m^2_\tau}\right)^{-1}\frac{R_{\rm IB}(s)}{S_{\rm EW}}\,,\label{eq:sf}
\end{eqnarray}
with
\begin{equation}
R_{\rm IB}(s)=\frac{{\rm FSR}(s)}{G_{\rm EM}(s)}\frac{\beta^3_0(s)}{\beta^3_-(s)}\left|\frac{F_0(s)}{F_-(s)}\right|^2\,.\label{eq:ib}
\end{equation}
In Eq.\,(\ref{eq:sf}), $m_\tau$ is the $\tau$ mass, $|V_{ud}|$ the CKM matrix element, ${\cal B}_{X^-}$ and ${\cal B}_e$ are the the branching fractions of $\tau^-\to X^-(\gamma)\nu_\tau$ (final state photon radiation is implied for $\tau$ branching fractions) and of $\tau^-\to e^-\bar{\nu}_e\nu_\tau$, $(1/N_x)dN_x/ds$ is the normalised invariant mass spectrum of the hadronic final state, and $R_{\rm IB}$ represents all the $s$-dependent isospin-breaking (IB) corrections and $S_{\rm EW}$ corresponds to short-distance electroweak radiative effects~\cite{bcvc09}.

\begin{figure*}[htbp]
\centering
\includegraphics[width=\columnwidth]{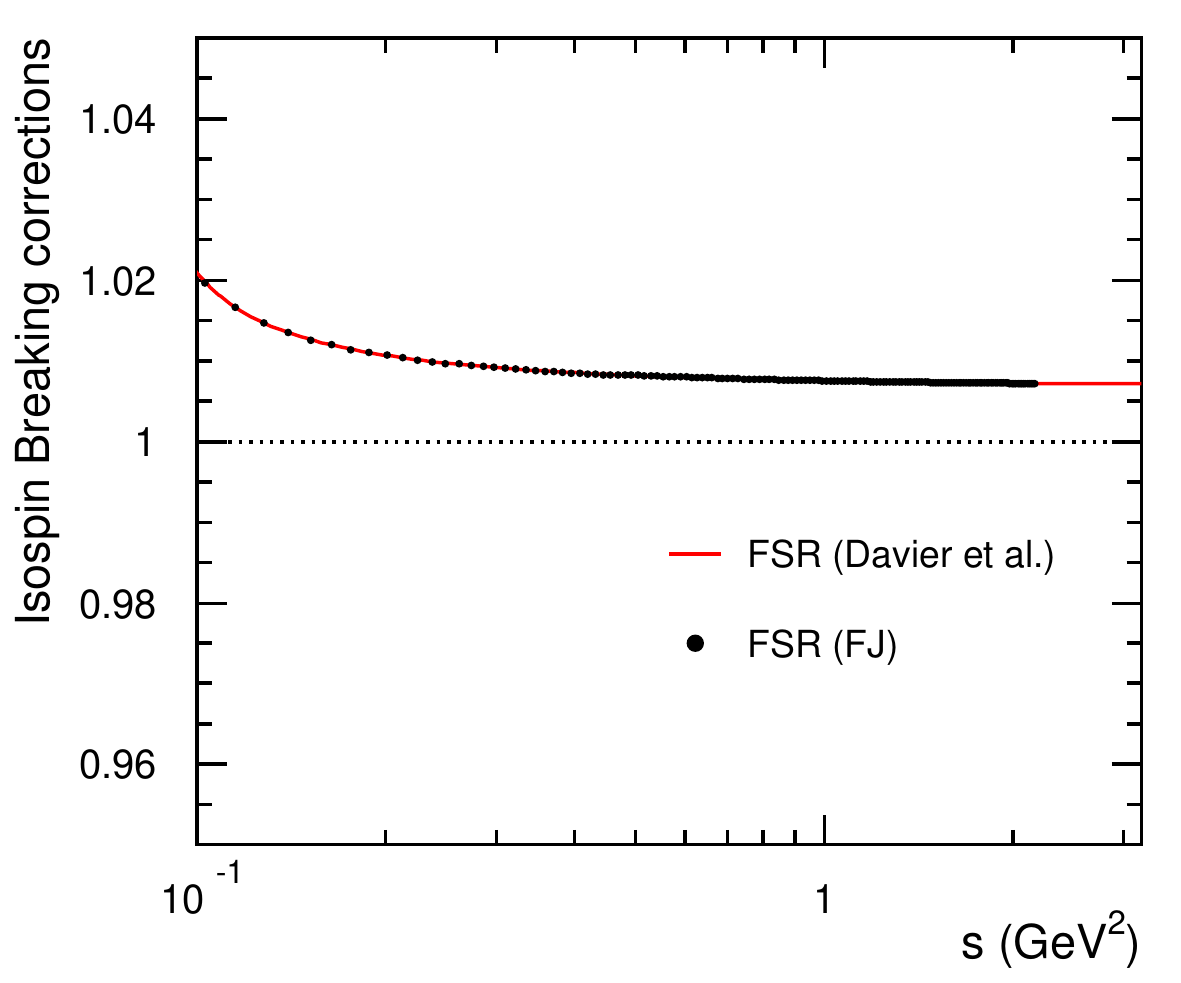}
\includegraphics[width=\columnwidth]{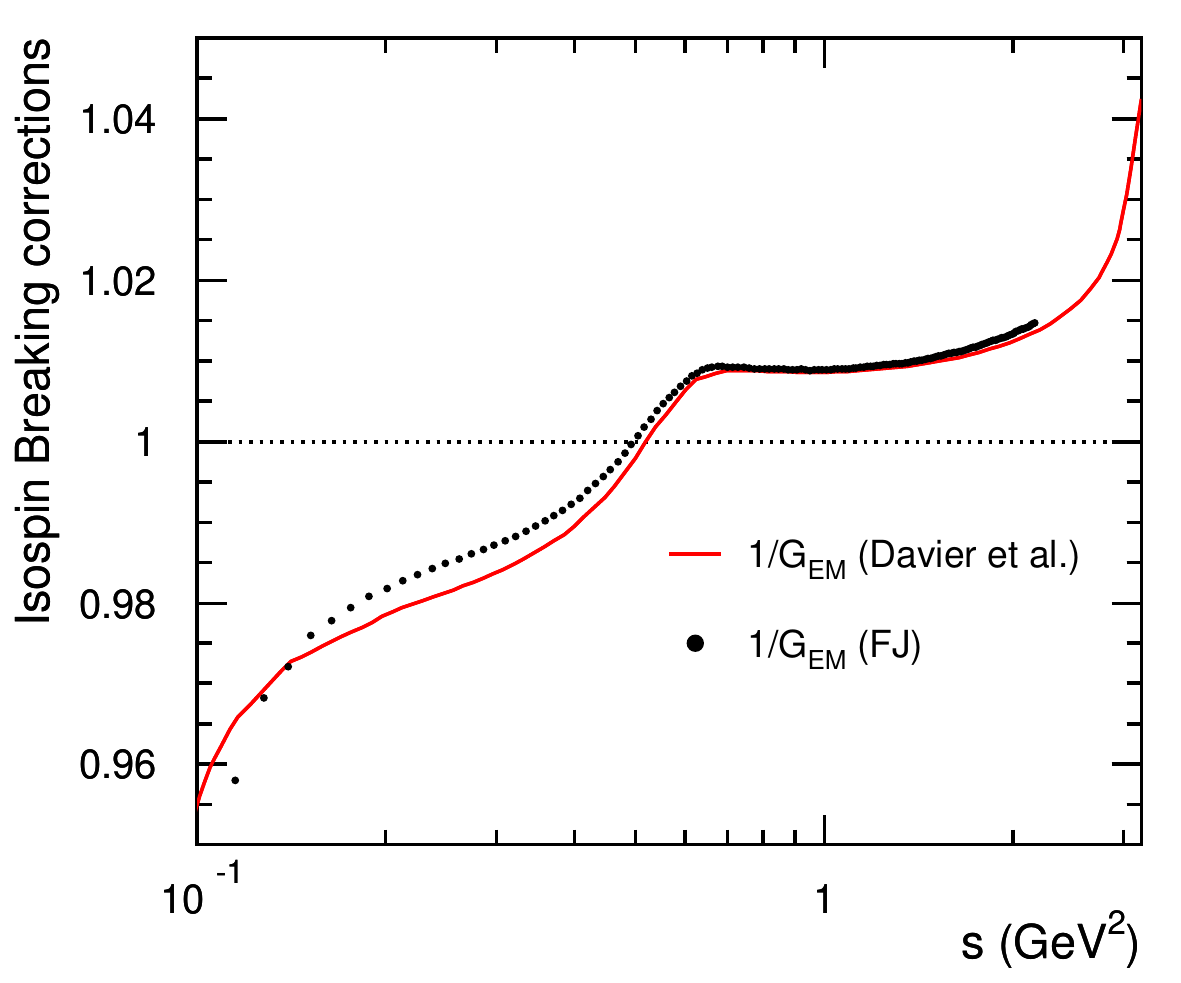}
\includegraphics[width=\columnwidth]{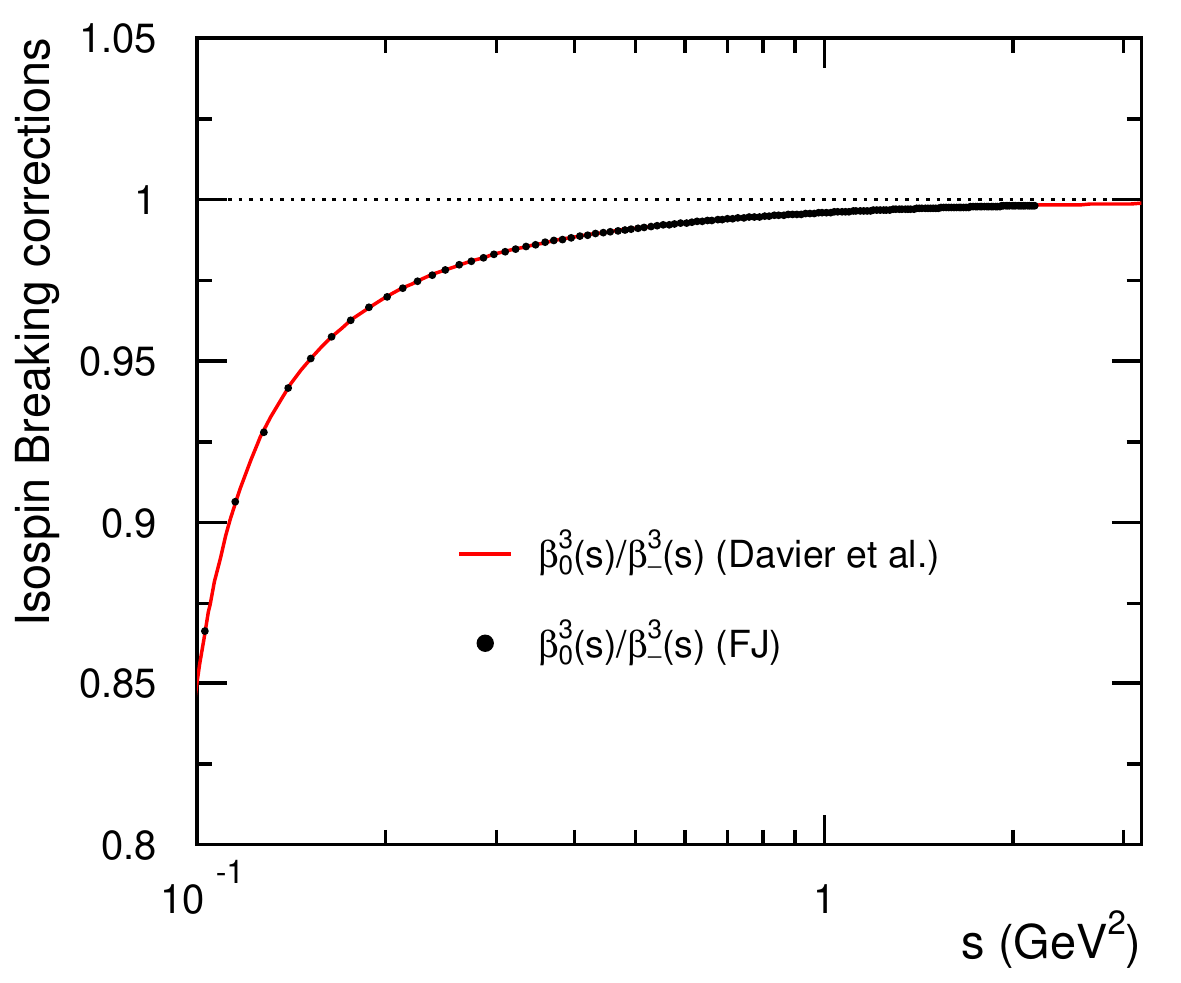}
\includegraphics[width=\columnwidth]{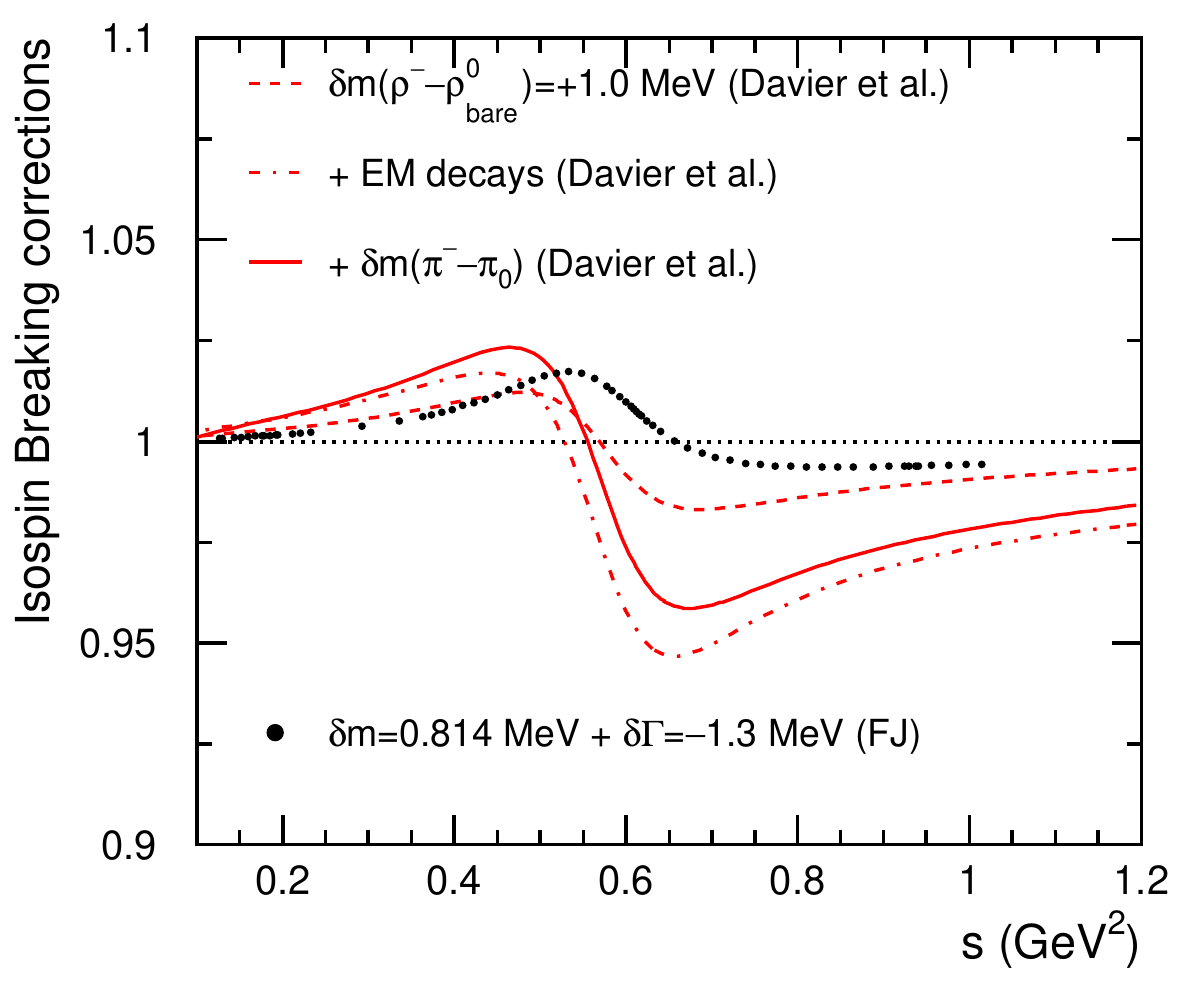}
\includegraphics[width=\columnwidth]{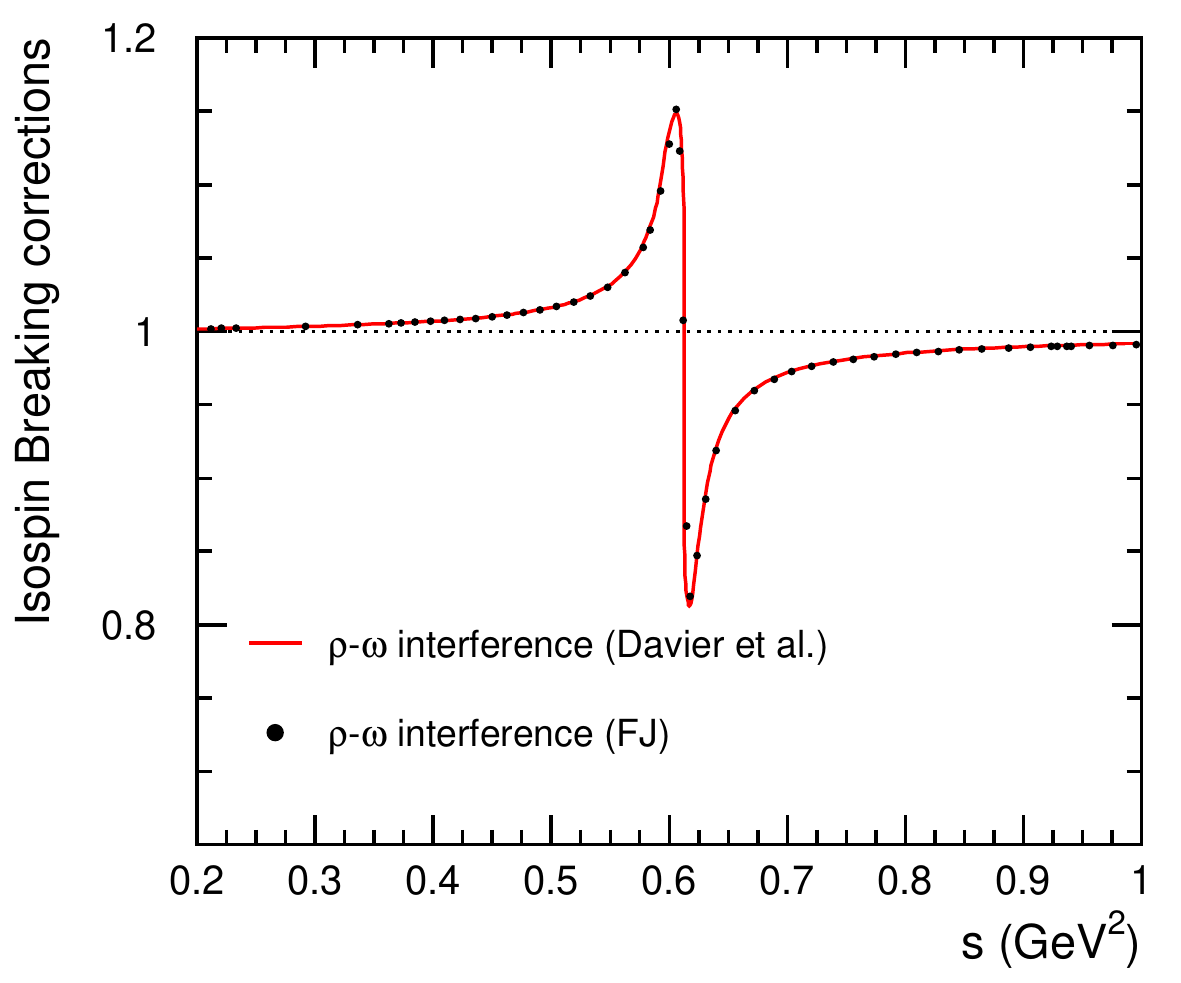}
\includegraphics[width=\columnwidth]{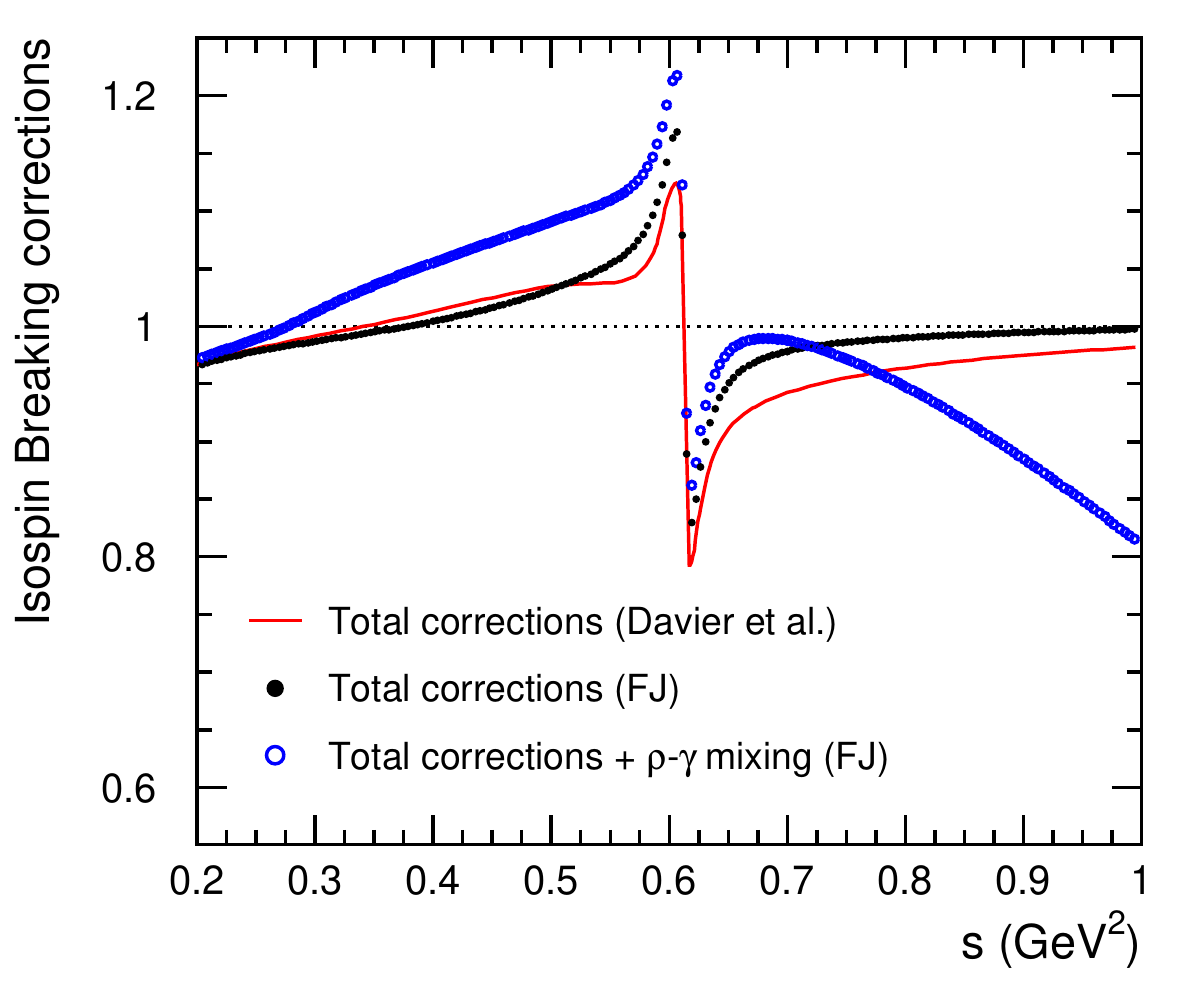}
\caption{Comparison of isospin-breaking corrections versus $s$ used by Davier et al.~\cite{bcvc09} and by JS~\cite{fred}. The different plots correspond to FSR (top left),  $1/G_{\rm EM}$ (top right), $\beta^3_0/\beta^3_-$ ratio term (middle left), the effect of the $\rho$ mass and width difference in the $|F_0/F_-|^2$ term (middle right), the effect of the $\rho-\omega$ interference in the $|F_0/F_-|^2$ term (bottom left) and the total corrections (bottom right). The difference between the open blue points and the solid black one in the last plot stems from the $\rho-\gamma$ mixing corrections proposed in~\cite{js11}}
\label{fig:ib}
\end{figure*}

The first term in Eq.\,(\ref{eq:ib}) is the ratio ${\rm FSR}(s)/G_{\rm EM}(s)$, where ${\rm FSR}(s)$ refers to the final state radiative corrections~\cite{JSS} in the $\pi^+\pi^-$ channel, and $G_{\rm EM}(s)$ denotes the long-distance radiative corrections of order $\alpha$ to the photon-inclusive $\tau^-\to \pi^-\pi^0\nu_\tau$ spectrum~\cite{bcvc09}. The energy dependent corrections used in~\cite{bcvc09} are compared in Fig.~\ref{fig:ib} with those by~\cite{fred}. The smaller correction of $G_{\rm EM}$ in~\cite{bcvc09} is due to the exclusion of the contributions involving the $\rho\omega\pi$ vertex.

The second correction term in Eq.\,(\ref{eq:ib}), $\beta^3_0(s)/\beta^3_-(s)$, arises from the $\pi^\pm-\pi^0$ mass splitting and is important only close the threshold (see Fig.~\ref{fig:ib}). 

The third IB correction term involves the ratio of the electromagnetic to weak form factors $|F_0(s)/F_-(s)|^2$ and is the most delicate one. Below 1\,GeV, the pion form factors are dominated by the $\rho$ meson resonance, such that IB effects mainly stem from the mass and width differences between the $\rho^\pm$ and $\rho^0$ mesons, and from $\rho^0-\omega$ mixing. The difference between the corrections used in~\cite{bcvc09} and those of~\cite{fred,js11} is mainly due to different width differences considered. The width difference $\delta\Gamma_\rho=\Gamma_{\rho^0}-\Gamma_{\rho^-}$ used in~\cite{bcvc09} was based on~\cite{flt07}
\begin{equation}
\delta\Gamma_\rho(s)=\frac{g^2_{\rho\pi\pi}\sqrt{s}}{48\pi}\left[\beta^3_0(s)(1+\delta_0)-\beta^3_-(s)(1+\delta_-)\right]\,,\label{eq:drho1}
\end{equation}
where $g_{\rho\pi\pi}$ is the strong coupling of the isospin-invariant $\rho\pi\pi$ vertex and $\delta_{0,-}$ denote radiative corrections for photon-inclusive $\rho\to \pi\pi$ decays, which include $\rho\to \pi\pi\gamma$. Contrary to expressions
\begin{equation}
\delta\Gamma_\rho=\frac{g^2_{\rho\pi\pi}}{48\pi}\left(\beta^3_0M_{\rho^0}-\beta^3_-M_{\rho^-}\right)\,,\label{eq:drho2}
\end{equation}
used in~\cite{js11}. The numerical values of Eqs.\,(\ref{eq:drho1}) at $M_\rho=775$\,MeV and (\ref{eq:drho2}) are +0.76\,MeV and $-1.3$\,MeV, respectively. Another small difference which contributes to the IB difference is from $\delta M_\rho=M_{\rho^-}-M_{\rho^0}$ of $1.0\pm 0.9$\,MeV~\cite{bcvc09} and 0.814\,MeV~\cite{js11}.

The effects of the IB corrections applied to $a_\mu^{\rm Had,LO}$ using $\tau$ data in the dominant $\pi\pi$ channel is shown in Table~\ref{tab:ib}~\cite{bcvc09} for the energy range between the $2\pi$ mass threshold and 1.8\,GeV. The first source corresponds to the effect of $S_{\rm EW}=1.0235\pm 0.003$~\cite{bcvc09}. The uncertainty of $G_{\rm EM}$ corresponds to the difference of two $G_{\rm EM}$ corrections shown in Fig.~\ref{fig:ib}. The quoted 10\% uncertainty on the FSR and $\pi\pi\gamma$ electromagnetic corrections is an estimate of the structure-dependent effects (pion form factor) in virtual corrections and of intermediate resonance contributions to real photon emission~\cite{bcvc09}. The systematic uncertainty assigned to the $\rho-\omega$ interference contribution accounts for the difference in $a_\mu^{\rm Had,LO}$ between two phenomenological fits, where the mass and width of the $\omega$ resonance are either left free to vary or fixed to their world average values. Some of the IB corrections depend on the form factor parametrisation used and the values quoted in Table~\ref{tab:ib}  corresponds to those of Gounaris-Sakurai (GS) parametrisation~\cite{gs} but the total uncertainty includes the full difference between the GS parametrisation and that of the K\"uhn-Santamaria (KS) parametrisation~\cite{bcvc09}. The total correction of $-16.07\pm 1.85$ is to be compared with the previous correction of $-13.8\pm 2.4$~\cite{dehz02} thus resulting a net change of $-6.9$.
\begin{table}[htb]
\centering
  \caption[.]{\label{tab:ib}
    Contributions to $a_\mu^{\rm Had,LO}~[\pi\pi, \tau]$ ($\times10^{-10}$) and ${\cal B}^{\rm CVC}_{\pi^-\pi^0}$ ($\times 10^{-2}$)
    from the isospin-breaking corrections.
    Corrections shown correspond to the 
    Gounaris-Sakurai (GS) parametrisation~\cite{bcvc09}. The total uncertainty includes the difference with the K\"uhn-Santamaria (KS) parametrisation quoted as $\delta({\rm GS}-{\rm KS})$} 
\small
\begin{tabular}{lcc}
\hline\noalign{\smallskip}
Source      & $\Delta a_\mu^{\rm Had,LO}[\pi\pi, \tau]$ &  $\Delta{\cal B}^{\rm CVC}_{\pi^-\pi^0}$ \\
\noalign{\smallskip}\hline\noalign{\smallskip}
$S_{\rm EW}$ &  $-12.21\pm0.15$ & $+0.57\pm 0.01$  \\
$G_{\rm EM}$  &  $ -1.92\pm0.90$ & $-0.07\pm 0.17$  \\
FSR                 &  $+4.67\pm0.47$ & $-0.19\pm 0.02$   \\
$\rho$--$\omega$ interference
                    & $+2.80\pm 0.19$ & $-0.01\pm 0.01$  \\
$m_{\pi^\pm}-m_{\pi^0}$ effect on $\sigma$
                    & $ -7.88$ & $+0.19$        \\
$m_{\pi^\pm}-m_{\pi^0}$ effect on $\Gamma_{\rho}$
                    & $+4.09$ & $-0.22$   \\
$m_{\rho^\pm}-m_{\rho^0_{\rm bare}}$ 
                    & $0.20^{+0.27}_{-0.19}$ & $+0.08\pm 0.08$    \\
$\pi\pi\gamma$, electrom. decays
                    & $ -5.91\pm0.59$ & $+0.34\pm 0.03$ \\
                    $\delta({\rm GS}-{\rm KS})$ & $-0.67$ & $-0.03$ \\
\noalign{\smallskip}\hline\noalign{\smallskip}
Total               & $-16.07\pm 1.85$ & $+0.69\pm 0.22$ \\
\noalign{\smallskip}\hline
\end{tabular}
\end{table}

In Table~\ref{tab:ib}, the effects of the IB corrections to the CVC prediction of ${\cal B}_{\pi\pi^0}$ are also shown. The prediction for the branching fraction of a heavy lepton decaying into a $G$-parity even hadronic final state, $X^-$,
\begin{eqnarray}
{\cal B}_X^{\rm CVC}&\!\!\!=\!\!\!&\frac{3}{2}\frac{{\cal B}_e|V_{ud}|^2}{\pi\alpha^2m^2_\tau}\int^{m^2_\tau}_{s_{\rm min}}ds s \sigma^I_{X^0}(s)\nonumber \\
&\!\!\!\times\!\!\!& \left(1-\frac{s}{m^2_\tau}\right)^2\left(1+\frac{2s}{m^2_\tau}\right)\frac{S_{\rm EW}}{R_{\rm IB}(s)}\,,
\end{eqnarray}
is derived from the vector current using the CVC relation with $s_{\rm min}$ being the threshold of the invariant mass-squared of the final state $X^0$ in $e^+e^-$ annihilation. CVC comparisons of $\tau$ branching fractions are of special interest because they are essentially insensitive to the shape of the $\tau$ spectral function, hence avoiding experimental difficulties, such as the mass dependence of the $\pi^0$ detection efficiency and feed-through, and biases from the unfolding of the raw mass distributions from acceptance and resolution effects.

\begin{figure}[htbp]
\centering
\includegraphics[width=\columnwidth]{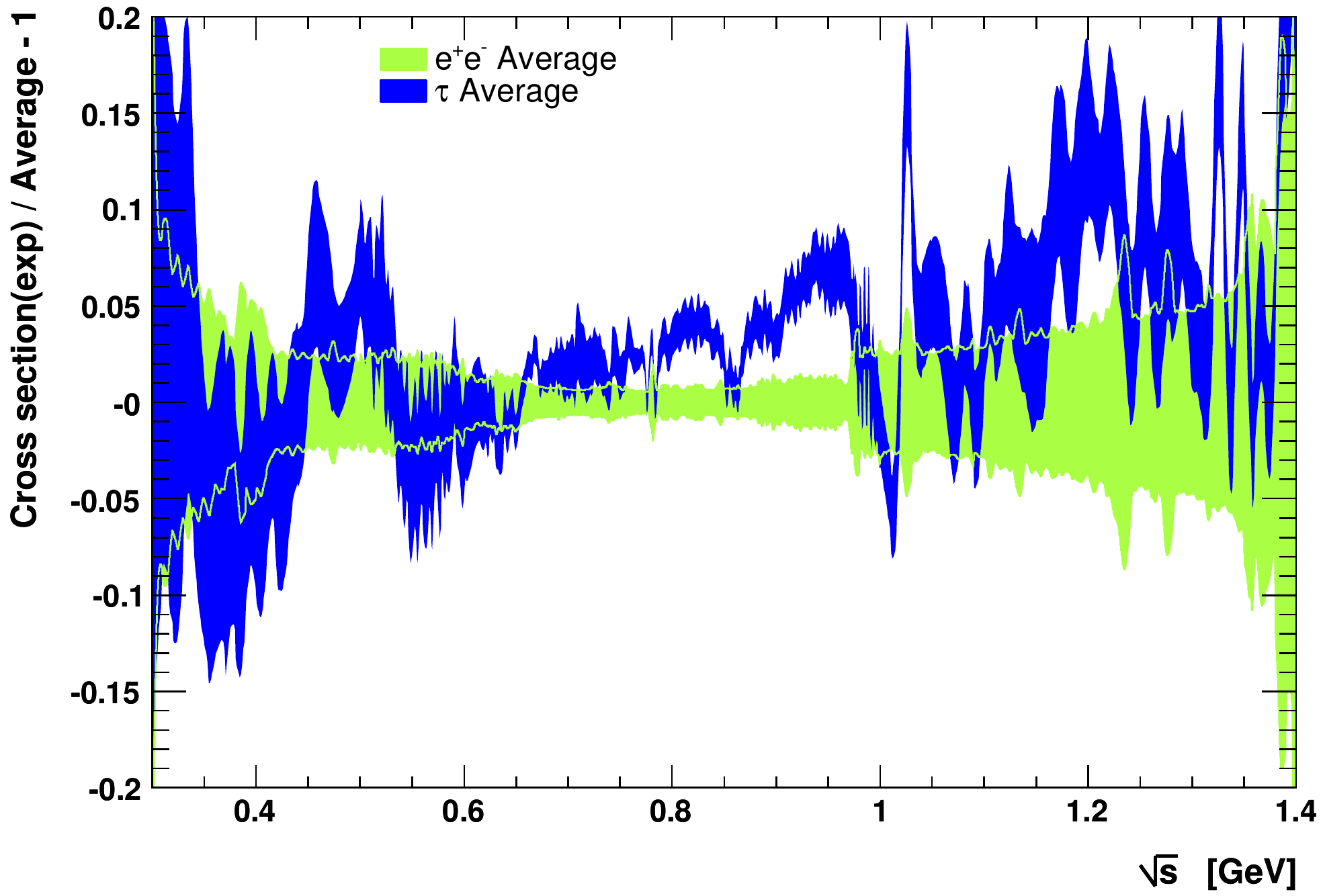}
\caption{Relative comparison between the combined $\tau$ after the IB corrections and $e^+e^-$ spectral functions (figure taken from~\cite{dhmyz09})}
\label{fig:chkib1}
\end{figure}

\begin{figure}[htbp]
\centering
\includegraphics[width=\columnwidth]{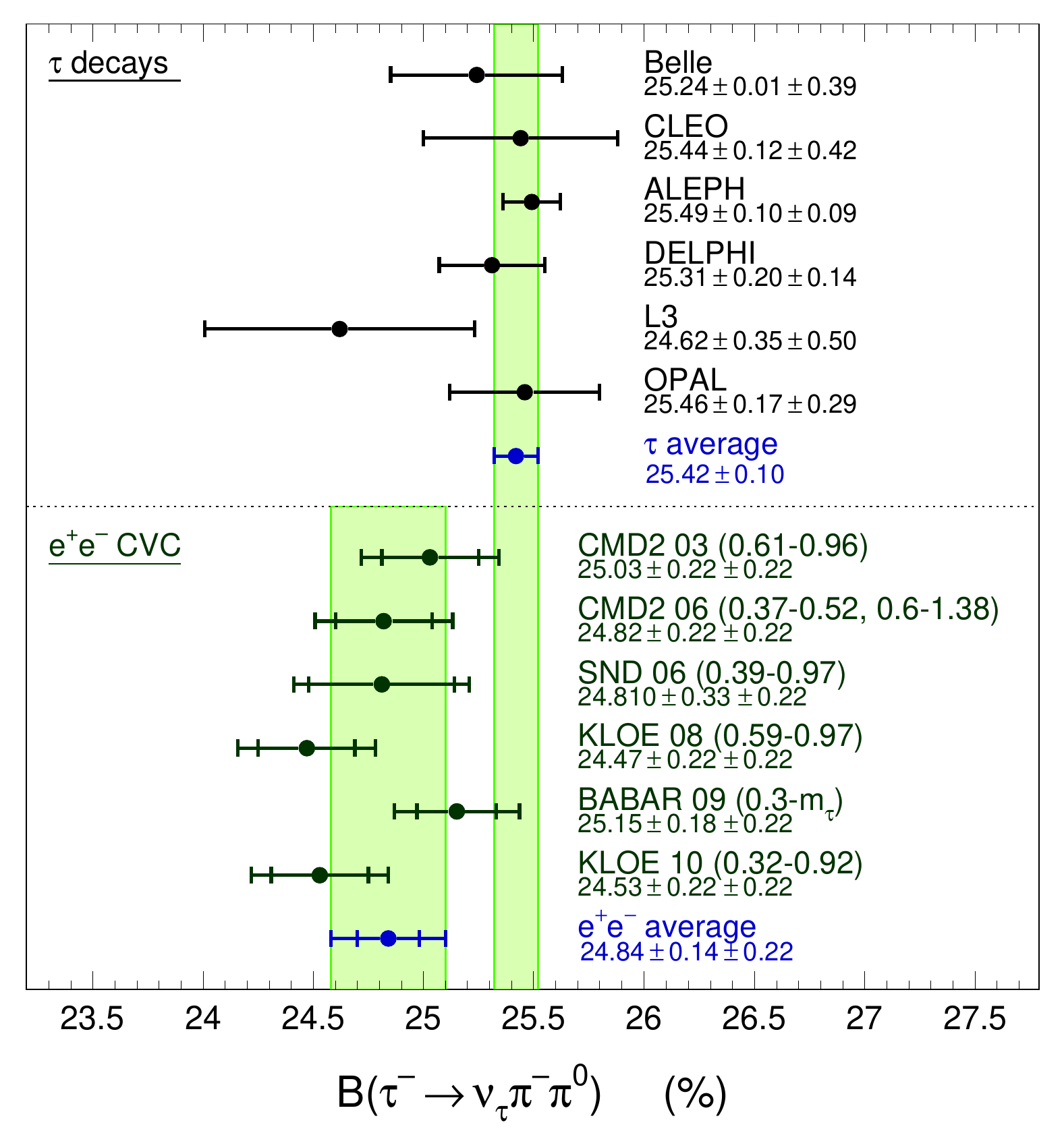}
\caption{The measured branching fractions for $\tau^-\to \pi^-\pi^0\nu_\tau$ compared to the predictions from the $e^+e^-\to \pi^+\pi^-$ spectral functions, applying the isospin-breaking corrections. The long and short vertical error bands correspond to the $\tau$ and $e^+e^-$ averages}
\label{fig:chkib2}
\end{figure}

Despite the improved IB corrections, there is still a sizeable difference between the $e^+e^-$ based prediction of $692.3\pm 4.2$ and the $\tau$ based one of $703.0\pm 4.4$~\cite{1312.7549,1312.1501}. The difference amounts to $10.7\pm 4.9$ corresponding to a deviation of 2.2\,$\sigma$. The shape of the combined $\tau$ spectral function after the IB corrections is also found different from the one from $e^+e^-$ data (Fig.~\ref{fig:chkib1}). The discrepancy is further reflected in the $\tau$ branching fractions (Fig.~\ref{fig:chkib2}). This used to be one of the open issues on the subject~\cite{1312.7549}.

\begin{figure*}[htbp]
\centering
\includegraphics[width=\columnwidth]{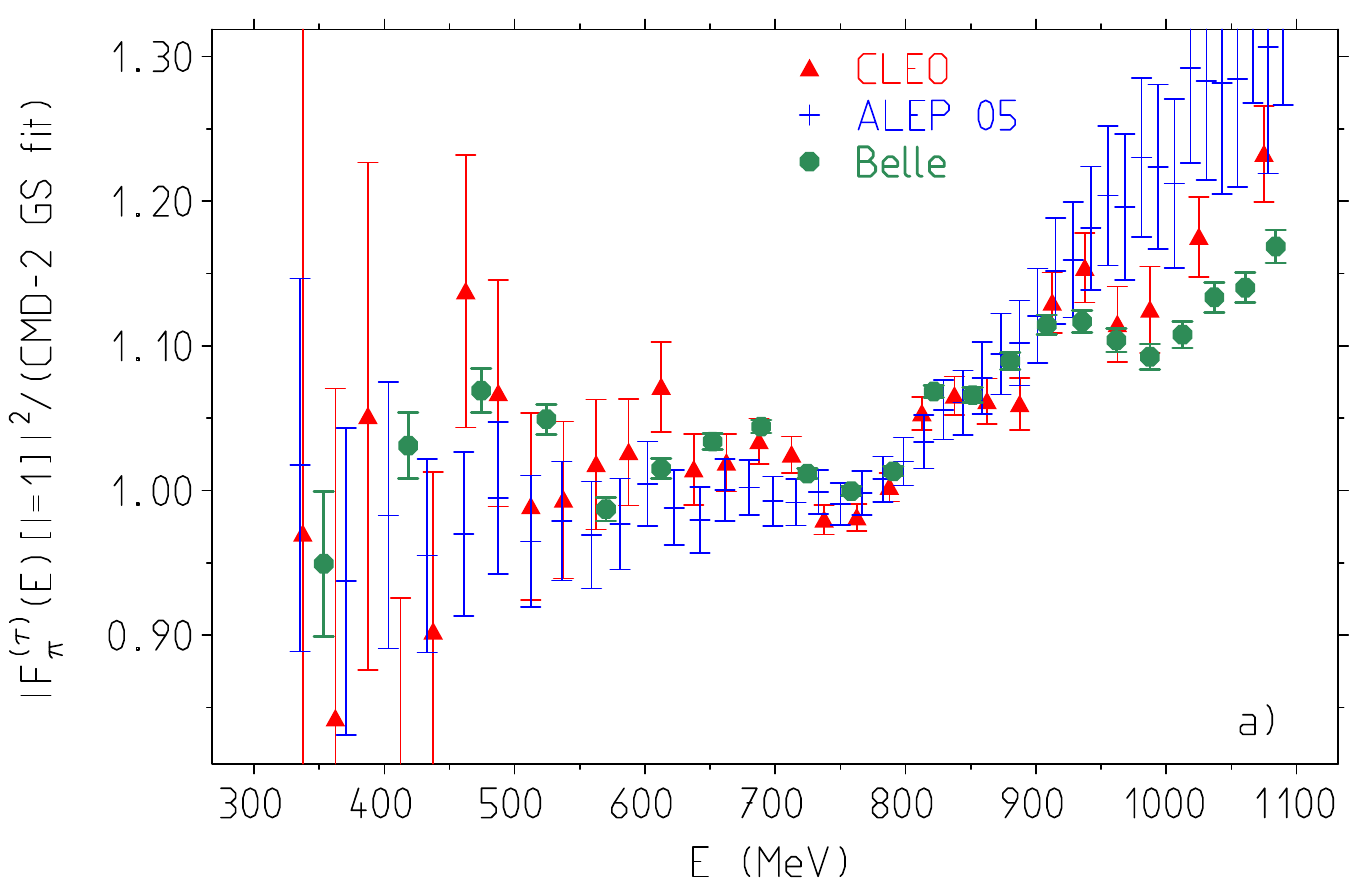}
\includegraphics[width=\columnwidth]{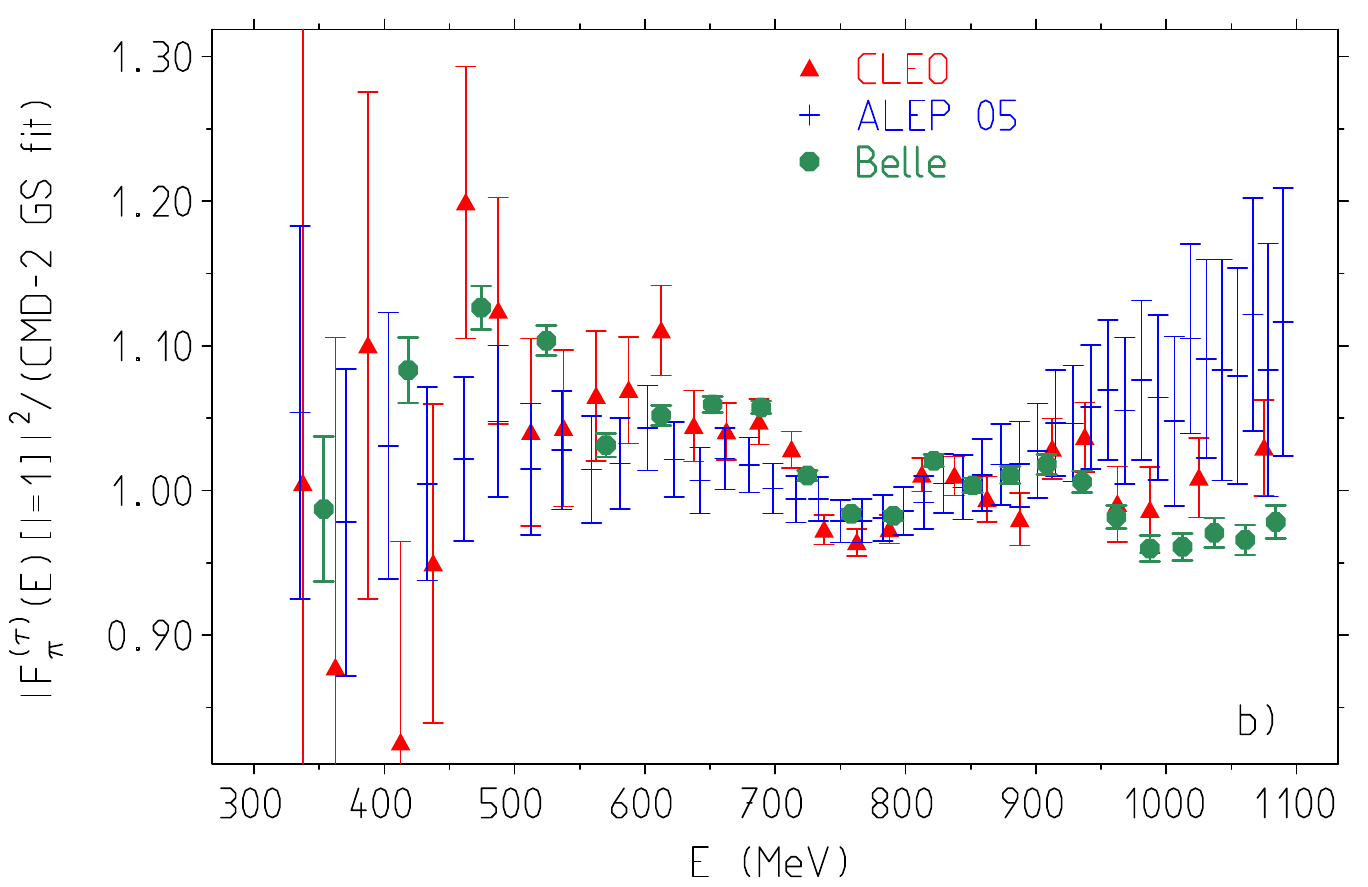}
\caption{$|F_\pi(E)|^2$ ratio $\tau$ versus $e^+e^- I=1$ (CMD-2 GS fit): left: uncorrected for $\rho-\gamma$ mixing; right: the same after correcting for it (figures taken from~\cite{js11})}
\label{fig:chkrgmixing} 
\end{figure*}

Recently, a model-dependent $\rho-\gamma$ mixing effect, which is absent in the $\tau$ data, was proposed in~\cite{js11} to explain the $e^+e^--\tau$ discrepancy. The proposed correction corresponds to the difference between the open blue points and the solid black points in Fig.~\ref{fig:ib} (bottom right).
The effect of the correction is shown in Fig.~\ref{fig:chkrgmixing}. After the correction, the $e^+e^--\tau$ difference above the rho peak looks indeed reduced, on the other hand, the agreement at the peak and below seems to become worse.

\section{Summary}\label{sec:summary}
The evaluation of the hadronic vacuum polarisation contributions to the muon magnetic anomaly $a_\mu$ (and the running of the QED coupling $\alpha(s)$) has a long history. The precision has been steadily improving thanks to more precise and complete cross section measurements of $e^+e^-$ annihilation into hadrons and of tau spectral functions on the one hand and the application of more advanced data interpolation and combination techniques in the dispersion relation approach on the other hand.

Some inconsistencies exist nevertheless among different $e^+e^-$ data sets in particular between Babar and KLOE which limit the accuracy of the combined results. There is also a discrepancy between the $e^+e^-$ data and the corresponding tau data after correcting for all known isospin-breaking effects. The $\rho-\gamma$ mixing effect is suggested to reduce the discrepancy. However, unlike for the analogous $\gamma-Z$ mixing, the correction here is model-dependent because of the $\rho$ hadronic structure.

The prospect is however good as improved or final measurements from Babar are expected for both the dominant $\pi\pi$ channel and the few other significant processes. In addition new data will be collected by CMD-3 and SND-2 at VEPP-2000, BES\,III at BEPC2, KLOE-2 at DA$\Phi$NE, and Belle\,II at superKEK-B in the next years. The new data and measurements should allow to improve the current precision of the leading-order hadronic vacuum polarisation contribution by a factor of at least two. Together with the expected improvement on the experimental side from Fermilab and J-PARC, the future of the muon magnetic anomaly will be very exciting.

\section*{Acknowledgment}
I am grateful for the fruitful collaboration with my colleagues and friends in particular M.~Davier, A.~Hoecker, G.~Lopez Castro and B.~Malaescu and I thank also F.~Jegerlehner for useful discussions on the isospin-breaking corrections.
%
%
%

\end{document}